\documentclass[aps,prl,final,twocolumn,superscriptaddress,%
               floatfix]{revtex4}       % Documentstyle for
\usepackage[latin1]{inputenc}                    % Codepage latin1
\usepackage{graphicx}                            % Graphics package
\usepackage{latexsym}                            % Load additional symbols
\usepackage{amsfonts}                            % Use AMS fonts
\usepackage{amssymb}                             % and symbols
\usepackage[mathscr]{eucal}                      % Euler font symbols
\usepackage{dcolumn}                             % Decimal-dot aligned
\usepackage{hyperref}                            % References as links

\graphicspath{{.}{graphics/}}                    % Path for graphics
\hypersetup{%
  debug=false,                                    % Don't be verbose
  a4paper=true,                                   % Use A4 paper
  pdfpagemode={UseOutline},                       % Open outline in
  pdftitle={The nucleon axial charge in full lattice QCD},                             % Title of publication 
  pdfauthor={R. G. Edwards, G. Fleming, Ph. Haegler, J.W. Negele, K. Orginos, A.V. Pochinsky,  D.B. Renner, D. G.  Richards, and
             W. Schroers},                        % Author(s)
  }

\begin{document}

% Setup title page
\preprint{MIT-@@@,DESY 05-203}
 \title[GA from the lattice]{The nucleon axial charge in full lattice QCD}
\author{ R.G.~Edwards} 
  \affiliation{Thomas Jefferson National Accelerator Facility, Newport News, VA 23606}
\author{G.T.~Fleming}
  \affiliation{Sloane Physics Laboratory, Yale University, New Haven, CT 06520}
\author{Ph.~H{\"a}gler}
  \affiliation{Department of Physics and
  Astronomy, Vrije Universiteit, 1081 HV Amsterdam, NL}
   \affiliation{Current affiliation: Institut f\"ur Theoretische Physik, TU M\"unchen, D-85747 Garching, Germany }
\author{J.W.~Negele} 
  \affiliation{Center for Theoretical Physics, Massachusetts Institute of Technology, Cambridge, MA 02139} 
\author{K.~Orginos}
  \affiliation{Center for Theoretical Physics, Massachusetts Institute of Technology,
  Cambridge, MA 02139} 
  \affiliation{Current affiliation: Department of Physics, College of William and Mary,
%P.O. Box 8795,
Williamsburg VA 23187-8795 \\
and Thomas Jefferson National Accelerator Facility, Newport News, VA 23606
}
\author{A.V.~Pochinsky} 
   \affiliation{Center for Theoretical Physics, Massachusetts Institute of Technology, Cambridge, MA 02139}
\author{D.B.~Renner} 
%  \affiliation{Center for Theoretical Physics, Massachusetts
%  Institute of Technology, Cambridge, MA 02139} 
  \affiliation{ Department of Physics, University of Arizona, 1118 E 4th
  Street, Tucson, AZ 85721} 
\author{D.G.~Richards}
  \affiliation{Thomas Jefferson National Accelerator Facility, Newport News, VA 23606}
\author{W.~Schroers} 
%  \affiliation{Center for Theoretical Physics, Massachusetts
%  Institute of Technology, 
%  Cambridge, MA 02139} 
  \affiliation{John von Neumann-Institut f\"ur Computing NIC/DESY, D-15738 Zeuthen, Germany}
  \collaboration{LHPC Collaboration}
   \date{\today}

\begin{abstract}
The nucleon axial charge is calculated as a function of the pion mass in full QCD. Using domain wall valence quarks and improved staggered sea quarks, we present the first calculation with pion masses as light as 354 MeV and volumes as large as (3.5 fm)$^3$. We show that finite volume effects are small for our volumes and that a constrained fit based on finite volume chiral perturbation theory agrees with experiment within 7\% statistical errors.
\end{abstract}

\maketitle

% --------------------------------------------------------------------------
%
%  Introduction
%
% --------------------------------------------------------------------------

\section{\label{sec:Introduction}Introduction}

\vspace{-.1cm}

 The nucleon axial charge,  $g_A $,  is a fundamental property of the nucleon, governing $\beta$ decay and providing a quantitative measure of spontaneous chiral symmetry breaking in low energy hadronic physics.  The axial vector form factor, $ g_A(q^2) $, is defined by the momentum space matrix element of the isovector axial current, $\vec A_\mu = \bar q \gamma_\mu \gamma_5 \vec \tau q $:
% \begin{equation}
%\langle N(p+q)| \vec A_\mu(q) | N(p) \rangle =  \bar u(p+q) \frac{\vec \tau}{2}[g_A(q^2) \gamma_\mu \gamma_5 + g_P(q^2)q_\mu \gamma_5] u(p)
%\end{equation}
%
\begin{eqnarray}
\langle N(p+q, s' )| \vec A_\mu(q) | N(p, s) \rangle =   \nonumber \hspace{3.0 cm} \\ 
\hspace{0 cm} \bar u(p+q, s') \frac{\vec \tau}{2}[g_A(q^2) \gamma_\mu \gamma_5 + g_P(q^2)q_\mu \gamma_5] u(p, s).
\end{eqnarray}
The axial charge is the form factor at zero momentum transfer and is accurately measured from neutron $ \beta$ decay,  $g_A \equiv g_A(0) = 1.2695 \pm 0.0029 $ (in units of the vector charge, $g_V$).
It is also useful to write the axial charge in terms of the zeroth moments with respect to the momentum fraction $x$ of the light cone quark helicity distributions, $\Delta q(x) = q_{\uparrow}(x) - q_{\downarrow}(x)$.  For each quark flavor $q$, these moments are defined and related to  forward  nucleon matrix elements of  twist-2 operators as 
%
%\begin{eqnarray}
%\langle 1 \rangle_{\Delta q} & \equiv & \int_0^1 dx [ \Delta q(x) + \Delta \bar q(x)] \nonumber \\
%&=& \frac{1}{2 M_N S_\mu} \langle N(p,s) | \bar q \gamma_\mu \gamma_5 q |N(p,s)\rangle,
%\label{eq:moment}
%\end{eqnarray}
%%
%and the axial charge is given by $g_A = \langle 1 \rangle _{\Delta u} -
% \langle 1 \rangle _{\Delta d}$.  
 \begin{eqnarray}
\langle 1 \rangle_{\Delta q} & \equiv & \int_0^1 dx [ \Delta q(x) + \Delta \bar q(x)] \nonumber \\
2 M_N S_\mu\langle 1 \rangle_{\Delta q} &=&  \langle N(p,s) | \bar q \gamma_\mu \gamma_5 q |N(p,s)\rangle,
\label{eq:moment}
\end{eqnarray}
and the axial charge is given by $g_A = \langle 1 \rangle _{\Delta u} -
 \langle 1 \rangle _{\Delta d}$.

In addition to governing neutron $\beta$ decay, the axial charge is physically important for several other reasons. The Adler-Weisberger sum 
rule\cite{Adler:1965ka,Weisberger:1965hp}  relates the excess of $g_A^2$ beyond the trivial value of 1 for a point, structureless nucleon to an integral over the difference between the $\pi^+ p$ and $\pi^- p$ scattering cross sections, which is dominated by coupling to  the $\Delta(1230)$ resonance.  The Goldberger-Treiman relation\cite{Goldberger:1958tr},  $g_A = f_\pi g_{\pi N N} / M_N$, relates it to the pion decay constant, pion nucleon coupling constant, and nucleon mass to an accuracy of five percent.  Finally, the matrix elements in Eq.~(\ref{eq:moment}) directly measure the contribution of each flavor to the spin of the nucleon, and the flavor singlet combination $\Sigma =   \langle 1 \rangle _{\Delta u} +  \langle 1 \rangle _{\Delta d}+  \langle 1 \rangle _{\Delta s}$ specifies the fraction of the total nucleon spin arising from the spin of the quarks.  Thus, a first principles calculation of the axial charge is an essential step in understanding key issues in hadron structure ranging from chiral dynamics to the spin content of the nucleon.

The axial charge is the ideal starting point in the quest for precision lattice calculation 
of hadron structure for several reasons.  It is accurately measured experimentally and the isovector combination $ \langle 1 \rangle_{\Delta u} - \langle 1 \rangle_{\Delta d} $ has no contributions from disconnected diagrams,  which are much more computationally demanding than the connected diagrams considered in this work. The functional dependence on both $m_\pi^2$ and volume is known at small  masses from chiral perturbation theory ($\chi$PT)\cite{Beane:2004rf,Detmold:2005pt}  and renormalization of the lattice axial vector current can be performed accurately nonperturbatively using the five dimensional conserved current for domain wall fermions. Thus, conceptually, it is a ``gold plated'' test of our ability to calculate hadron observables from first principles on the lattice.  In addition, since it is known to be particularly sensitive to finite lattice volume effects that reduce the contributions of the pion cloud\cite{Sasaki:2001th,Khan:2004vw},  it is also a stringent test of our control of finite volume artifacts. 

% --------------------------------------------------------------------------
%
% Lattice Calculation in Chiral Regime
%
% --------------------------------------------------------------------------
\vspace{-.1cm}

\section{\label{sec:Lattice Calculation}Lattice Calculation in the Chiral Regime}

\vspace{-.1cm}

The computational challenge of precision lattice QCD lies in simultaneously approaching the continuum limit of small lattice spacing, the chiral limit of low quark masses, and the large volume limit.  For nucleon structure, where the pion cloud extends to large distances and plays a major role, the latter two requirements are particularly severe.  Instead of referring to the unobservable bare quark mass, since $m_q \propto m_\pi^2$, it is convenient to express the quark mass dependence of observables by their dependence on $m_\pi^2$.   To include the essential physics of the pion cloud, the box size must be large compared to the pion Compton wavelength, $m_\pi^{-1}$, and  the ultimate computational cost of a full QCD lattice calculation in such a box including dynamical sea fermions has mass dependence approximately $m_\pi^{-9}$.  

To address this exceedingly demanding regime of light pion mass and large volume with current computational resources, we have used a hybrid combination of computationally economical  staggered sea quark configurations generated using the so-called Asqtad improved action by the MILC collaboration\cite{Orginos:1999cr,Orginos:1998ue} with domain wall valence quarks that have chiral symmetry on the lattice. Improved staggered quarks have been strikingly successful in precision calculations\cite{Davies:2003ik} and even predictions of heavy quark systems\cite{Kronfeld:2005fy}.  Although there is not yet a proof that the continuum limit of the fourth root of the staggered determinant has the correct continuum limit, recent renormalization group arguments suggest that it does\cite{Shamir:2005sv}. Assuming that the staggered continuum limit is correct, the continuum limit of the hybrid calculation will also be correct. Both the Asqtad and domain wall actions have leading errors of order $a^2$ in the lattice spacing
$a$.

\begin{table}[t]
\caption{The number of configurations, pion mass, renormalization factor, and axial charge for each of the measurements on two lattice volumes.}
\label{tab:latticedata}
\begin{ruledtabular}
\begin{tabular}{@{}*{4}{l}}
\hline\\[-.32cm]
 \multicolumn{4}{c}{ $20^3\times 32$ lattice   $ V =(2.5 \, {\rm fm} )^3 $} \\
 \hline
 $\#$ configs. & 
$m_{\pi}$ (MeV) & $Z_A$ &
$g_A $\\
 \hline
425 & 761 (2) & 1.1296 (6) & 1.167 (11) \\
350 & 693 (3) & 1.1197 (6) & 1.153 (16) \\
564 & 594 (2) & 1.1085 (5) & 1.193 (16) \\
486 & 498 (2) & 1.0994 (4) & 1.173 (29) \\
655 & 354 (3) & 1.0847 (6) & 1.244 (58) \\
\hline\\[-.32cm]
 \multicolumn{4}{c}{ $28^3\times 32$ lattice   $ V = (3.5 \, {\rm fm} )^3 $} \\
 \hline
 $\#$ configs. & 
$m_{\pi}$ (MeV) & $Z_A$ &
$g_A $\\
 \hline

270 & 353 (1) & 1.0840 (5) & 1.212 (59) \\

\hline
\end{tabular}
\end{ruledtabular}
\end{table}

As tabulated in Table~\ref{tab:latticedata}, calculations were performed at five pion masses down to  354 MeV on $20^3 \times 32$ lattices with a spatial volume $ V = (2.5\,{\rm  fm} )^3 $ and the lightest mass, which is most sensitive to finite volume effects, was also calculated on a  $28^3 \times 32$ lattice with a spatial volume $ V =(3.5 {\rm fm} )^3 $. The scale is set by the lattice spacing $ a = 0.12406$ fm determined from heavy quark spectroscopy\cite{Aubin:2004wf} with an uncertainty of 2\%.  To reduce the effect of dislocations at this lattice spacing and thus improve the chiral properties of the domain wall fermions, the fermion action was defined 
to include HYP smearing\cite{Hasenfratz:2001hp} of the gauge fields, in which each link is replaced by a sum of links within the unit hypercube. This HYP smearing changes the action by an irrelevant operator, is restricted in range to  a single lattice spacing, and has coefficients  defined to minimize fluctuations of Wilson loops.  

The domain wall fermion\cite{Kaplan:1992bt,Furman:1994ky} propagators are calculated on a 5-dimensional lattice with fifth dimension $L_5$ and 5-dimensional mass $M_5$. The physical quark fields, $q(\vec x, t)$, reside on the 4-dimensional boundaries with fifth coordinate 1 and $L_5$ and have bare quark parameter $(a m)^{DWF}_q$. The value $M_5 = 1.7$ was chosen  on the basis of spectral flow analyses to optimize the evaluation of domain wall propagators, and $L_5 = 16$ was chosen\cite{Renner:2004ck} to ensure that the residual mass characterizing residual chiral symmetry breaking is always less than 10\% of the physical quark mass for the pion masses in Table \ref{tab:latticedata}.  The values of $(a m)^{DWF}_q$ were tuned to reproduce the  lightest pion mass obtained with the Asqtad action\cite{Aubin:2004wf} and as shown in Ref.\cite{Renner:2004ck}, the resulting nucleon masses were also consistent with the Asqtad masses.

Lattice matrix elements of the axial current, $\langle N | \bar q \gamma_\mu \gamma_5 q | N \rangle^{lattice}$, are calculated from the ratio of three point functions to two point functions using the same smeared nucleon sources and methodology as in reference~\cite{Dolgov:2002zm}.  The renormalization factor, $Z_A$, for the four dimensional axial current operator $A_\mu =   \bar q \gamma_\mu \gamma_5 q $ is calculated using the five dimensional conserved axial current for domain wall fermions ${\cal A}_\mu$ by the relation\cite{Blum:2000kn} $ \langle {\cal A}_\mu(t) A_\mu(0)\rangle = Z_A  \langle A_\mu(t) A_\mu(0)\rangle$. 

% --------------------------------------------------------------------------
%
%  Lattice Results
%
% --------------------------------------------------------------------------

\section{\label{sec:Lattice Results}Lattice Results}

The lattice renormalization constants, $Z_A$, and values of the renormalized axial charge, $g_A = Z_A \langle N | \bar q \gamma_\mu \gamma_5 q | N \rangle^{lattice}$, are tabulated in Table \ref{tab:latticedata} for the five pion masses and two volumes considered in this work. Measurement of $g_A$ to 5 \% accuracy at $m_\pi $= 354 MeV is a major achievement of this work, requiring 655 configurations on the $20^3$ lattice and correspondingly fewer on the larger lattice where volume averaging improves statistics.

%{\bf Note: the chiral fit and discussion are subject to change}
%The pion mass dependence of $ g_A$ can be calculated from chiral perturbation theory\cite{Hemmert:2003cb,Beane:2004rf,Detmold:2002nf}, where at sufficiently low pion mass, the functional form follows from the symmetries of QCD and the coefficients are low energy constants of QCD that in principle can be calculated on the lattice. In this work, we use the form of Ref.~\cite{Hemmert:2003cb}, which involves six constants:  $f_\pi$ (the pion decay constant), $m_\Delta-m_N$ (the Delta - nucleon mass splitting), $g_A$, $g_{\Delta \Delta}$, and $g_{N\Delta}$ (the couplings of the axial current between two nucleons, two Deltas, or a  nucleon and Delta, respectively), and a counterterm $C$.  

The dependence of $ g_A$ on the pion mass\cite{Hemmert:2003cb,Detmold:2002nf,Beane:2004rf,Detmold:2005pt}, and on both the mass and volume\cite{Beane:2004rf,Detmold:2005pt} is known from $\chi$PT, where at sufficiently low pion mass, the functional form follows from the symmetries of QCD and the coefficients are low energy constants of QCD that in principle can be calculated on the lattice. We have used the  $\chi$PT form of Ref.~\cite{Beane:2004rf}  and verified that it   agrees closely numerically with the other corresponding infinite volume\cite{Hemmert:2003cb,Detmold:2005pt} and finite volume\cite{Detmold:2005pt} results for our masses and volumes.
Our  $\chi$PT fit includes six constants:  $f_\pi$ (the pion decay constant), $m_\Delta-m_N$ (the Delta - nucleon mass splitting), $g_A$, $g_{\Delta \Delta}$, and $g_{N\Delta}$ (the couplings of the axial current between two nucleons, two Deltas, or a  nucleon and Delta, respectively), and a counterterm $C$.

\begin{figure}[t]
\begin{center}
 \vspace*{-1.3cm}
 \hspace*{-0.4 cm}  
 \hspace*{0cm} \raisebox{0cm}{\includegraphics[scale=0.35,clip=true,angle=-90]{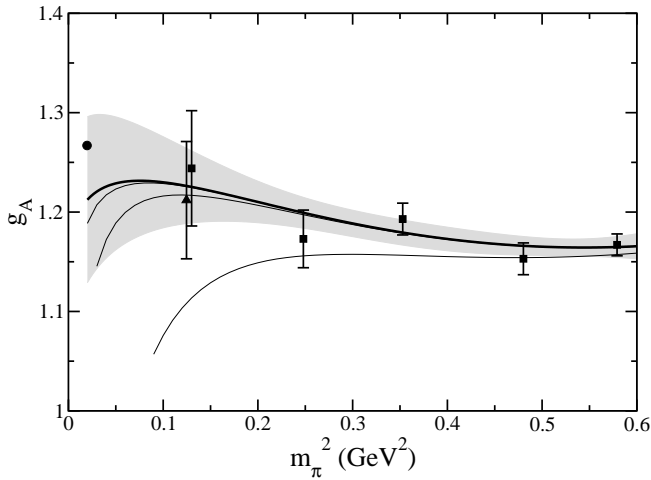}}
%  \vspace*{0cm}
%  
  \caption{\label{fig:gAfit}Nucleon axial charge $g_A$ as a function of
  the pion mass. Lattice data are denoted by squares (smaller volume) and a triangle (larger volume), the lowest smaller volume point is displaced slightly to the right for clarity, and experiment is denoted by the circle.  The heavy solid line and shaded error band show the $\chi$PT fit to the finite volume data evaluated in the infinite volume limit, and the lines below it show the behavior of this chiral fit in boxes of finite volume $L^3$, as $L$ is reduced to 3.5, 2.5, and 1.6 fm respectively.} 
 \end{center}
 \end{figure}

%Figure \ref{fig:gAfit} shows the lattice data and a fit to it from chiral perturbation theory.  In the absence of  lattice calculations of $g_A$ at still lower pion masses, it is not presently possible to do a complete extrapolation from lattice measurements alone.  Hence, following Ref.~\cite{Hemmert:2003cb}, we perform a constrained fit and the solid curve is determined by setting $f_\pi$, $m_\Delta-m_N$, and $g_{N\Delta}$ to their physical values\cite{Hemmert:2003cb} and performing a least squares fit for   $g_A$, $g_{\Delta \Delta}$, and $C$.  The error band arising from this three parameter fit is shown in Figure \ref{fig:gAfit} and the resulting value for the axial charge at the physical pion mass is $g_A (m_\pi = 140 \,{\rm MeV}) = 1.21 \pm 0.07$.  Given the  smooth behavior of the chiral fit and the small magnitude of the chiral logs, it is clear that the extrapolation for the axial charge is quite benign, and that the lattice data is extrapolating convincingly towards experiment. 
Figure~\ref{fig:gAfit} shows the lattice data and our fit to it using finite volume $\chi$PT.  The  $\chi$PT function  was  fit to each data point at the corresponding mass and finite volume, and the parameters of the fit were then used to determine the infinite volume axial charge.  In the absence of  lattice calculations of $g_A$ at still lower pion masses, it is not presently possible to do a complete extrapolation from lattice measurements alone.  Hence, following Ref.~\cite{Hemmert:2003cb}, we performed a constrained fit and the heavy solid curve is determined by setting $f_\pi$, $m_\Delta-m_N$, and $g_{N\Delta}$ to their physical values\cite{Hemmert:2003cb} and performing a least squares fit for   $g_A$, $g_{\Delta \Delta}$, and $C$.  The error band arising from this three parameter fit is shown in Fig.~\ref{fig:gAfit} and the resulting value for the axial charge at the physical pion mass is $g_A (m_\pi = 140 \,{\rm MeV}) = 1.212 \pm 0.084$. Given the  smooth behavior of the chiral fit and the small magnitude of the chiral logs, it is clear that the extrapolation for the axial charge is quite benign, and that the lattice data is extrapolating convincingly towards experiment. 

%Accurate calculation of hadron structure requires that the lattice volume be sufficiently large to contain the full contribution of the pion cloud. Although there is no danger of missing part of the axial charge on a periodic lattice\cite{Jaffe:2001eb}, both the Adler Weisberger sum rule\cite{Adler:1965ka,Weisberger:1965hp} and the prominent role of pion-Delta excitations in chiral perturbation theory\cite{Hemmert:2003cb,Beane:2004rf} suggest that $g_A$ may be particularly sensitive to finite volume effects.  Quenched calculations~\cite{Sasaki:2001th} have shown that increasing the box length from 1.2 fm to 2.4 fm increases $g_A$ by the order of 10\% for pion masses ranging from 550 to 870 MeV, and unquenched calculations for 770 MeV pions\cite{Khan:2004vw} have shown that increasing $L$ from 1.1 to 2.2 fm increases $g_A$ by  20\%. Hence, it is important to observe that at the lightest quark mass of 354 MeV, our results on lattices with $L$ = 2.5 and 3.5 fm in Table \ref{tab:latticedata} are statistically consistent.  Furthermore, this behavior is also consistent with the prediction of chiral perturbation theory\cite{Beane:2004rf} that for 300 MeV pions, the difference in $g_A$ for our two volumes  should be of the order of 0.02.  

Although there has been concern that $g_A$ is particularly sensitive to finite volume effects\cite{Jaffe:2001eb,Sasaki:2001th,Khan:2004vw}, Fig.~\ref{fig:gAfit} shows these effects are well under control and introduce negligible errors for our volumes. The light curves show the behavior expected from  $\chi$PT  in volumes $L^3$ with L of 3.5, 2.5, and 1.6 fm.  Note that at the lightest mass, our 3.5 and 2.5 fm results are statistically indistinguishable, consistent with the $\chi$PT change of less than 1 \%, and that the corrections applied in correcting our data from 2.5 or 3.5 fm to infinity in the $\chi$PT fit are quite small.  At heavier masses, although the truncated $\chi$PT expansion is not quantitatively reliable, the finite volume effects are physically suppressed. The order of magnitude of the corrections from 1.6 to 2.5 fm is also consistent with the fact that quenched calculations\cite{Sasaki:2001th} have shown that increasing the box length from 1.2 fm to 2.4 fm increases $g_A$ by the order of 10\% for pion masses ranging from 550 to 870 MeV, 
and unquenched calculations for 770 MeV pions\cite{Khan:2004vw} have shown that increasing $L$ from 1.1 to 2.2 fm increases $g_A$ by  20\%.

%In addition to the statistical error arising from fitting the parameters $g_A$, $g_{\Delta \Delta}$, and $C$, several systematic errors may be estimated. The three constrained parameters can be calculated directly on the lattice. 
%Calculation of  $f_{\pi}$ yields 92.4 MeV $\pm$ 3\%, corresponding to an error in $g_A$ of ??, and the current uncertainty of roughly 18\%  in $m_\Delta-m_N$ corresponds an error of ??. The 2 \% error in the lattice scale will induce a negligible effect since the curve in Fig.~\ref{fig:gAfit} is so flat. An alternative lattice renormalzation method based on calculating the ratio of the axial and vector charges, which should have the same renormalization constant in the chiral limit, yields discrepancies less than 2\% for the heaviest masses and statistically indistinguishable results at lighter masses, suggesting that the renormalization error is less than a few percent.  

%\begin{figure}[t]
%\begin{center}
% \vspace*{-1.0cm}
% \hspace*{-0.2 cm}  
% \includegraphics[scale=0.32,clip=true,angle=0]{g_A-lhpc.eps}
% \includegraphics[scale=0.35,clip=true,angle=-90]{g_A.fig2.for.david.ps}
\begin{figure}[t]
\begin{center}
 \vspace*{-1.3cm}
 \hspace*{-0.4 cm}  
 \hspace*{0cm} \raisebox{.0cm}{\includegraphics[scale=0.35,clip=true,angle=-90]{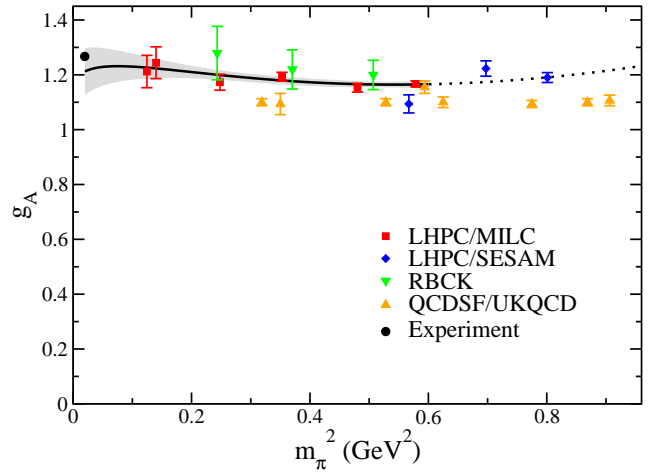}}
  \vspace*{-.15cm}
 \caption{Comparison of all full QCD calculations of  $g_A$, as described in the text. The solid line and error band denote the infinite volume $\chi$PT fit
of Fig.~1 and its continuation to higher masses is indicated by the dotted line.
 }
 \label{fig:gAall}
 \end{center}
%\vspace*{-.2cm}
\end{figure}

In addition to the statistical error arising from fitting the parameters $g_A$, $g_{\Delta \Delta}$, and $C$, several systematic errors may be estimated. The three constrained parameters can be calculated directly on the lattice and the linear response of our chiral fit to varying each of of them shows very weak dependence.
Calculation of  $f_{\pi}$ on our lattices yields 92.4 MeV $\pm$ 3\%, corresponding to an error in $g_A$ of 0.10\%, and a rough calculation of $m_\Delta-m_N$, (which can be improved) with  18\%  uncertainty corresponds to an error of 0.29\%.  Although we have not yet calculated $g_{N\Delta}$, it should be calculable to 20\%, corresponding to an error in $g_A$ of 0.13 \%. Thus the total error from the constrained parameters is much less than a percent. The  error in the lattice scale, which can shift all masses by 2\%, will induce a negligible effect since the curve in Fig.~\ref{fig:gAfit} is so flat. An alternative lattice renormalzation method based on calculating the ratio of the axial and vector charges, which should have the same renormalization constant in the chiral limit, yields discrepancies less than 2\% for the heaviest masses and statistically indistinguishable results at lighter masses, suggesting that the renormalization error is less than a few percent.

Finally, we compare our results with the world's supply of unquenched lattice calculations of $g_A$ in Fig.~\ref{fig:gAall}.  
The dotted line shows the continuation of our $\chi$PT fit to higher masses to guide the eye, but is not quantitatively reliable due to the limitations of $\chi$PT at such large pion masses. 
Our previous calculations\cite{Dolgov:2002zm} using SESAM configurations\cite{Eicker:1998sy} with dynamical Wilson quarks and spatial box size 1.5 fm are shown for the heaviest three quark masses.  These are consistent with the current hybrid calculations within statistics, although the lowest mass point may have a small systematic shift downwards due to the small volume.  The three points by the RBCK collaboration\cite{Ohta:2004mg} using dynamical domain wall fermions in a 1.9 fm box are also consistent with our present results within statistics. The initial results by the QCDSF/UKQCD collaboration\cite{Khan:2004vw} using improved Wilson dynamical fermions in boxes ranging from 1.5 to 2.2 fm are systematically lower than the results of the other three calculations, although finite volume effects and differences in renormalization away from the chiral limit may account for the discrepancy.

\section{\label{sec:Conclusions}Conclusions}

In summary, we have calculated $g_A$ in full QCD in the chiral regime.  The hybrid combination of improved staggered sea quarks and domain wall valence quarks enabled us to extend calculations to the lightest mass, 354 MeV, and largest box size, 3.5 fm, yet attained, and to obtain statistical accuracy of 5\% with negligible error from volume dependence. Chiral perturbation theory implies mild  dependence on the  pion mass, and a three parameter constrained fit  yields an excellent fit to the data and generates an error band of size 7\% at the physical pion mass which overlaps experiment. Thus, this calculation represents a significant milestone in the quest to calculate hadron structure from first principles.  Building on this result, future challenges will include extension of the range of pion masses to  include 300 and 250 MeV to enable an unrestricted determination of the parameters of $\chi$PT from lattice data, generalization to partially quenched calculations and the application of partially quenched hybrid $\chi$PT\cite{Bar:2005tu}, and evaluation of disconnected diagrams to enable calculation of the quark spin structure of the nucleon.

% --------------------------------------------------------------------------
%
%  Acknowledgments
%
% --------------------------------------------------------------------------

\begin{acknowledgments}
 
 We are grateful for helpful discussions with Will Detmold, Martin Savage,  Tony Thomas, Wolfram Weise, and Ross Young concerning chiral extrapolation. 
This work was supported by the DOE Office of Nuclear Physics under contracts DE-FC02-94ER40818, DE-FG02-92ER40676, and DE-AC05-84ER40150, the EU Integrated Infrastructure Initiative Hadron Physics (I3HP)
under contract RII3-CT-2004-506078 and by the DFG under contract
FOR 465 (Forschergruppe Gitter-Hadronen-Ph\"anomenologie).  Computations were performed on clusters at Jefferson Laboratory and at ORNL using time awarded under the SciDAC initiative. We are indebted to members of the MILC  and SESAM collaborations for providing the dynamical quark configurations which made our full QCD calculations possible.

\end{acknowledgments}

% --------------------------------------------------------------------------
%
%  Bibliography
%
% --------------------------------------------------------------------------

\bibliography{gA_references1}

% --------------------------------------------------------------------------
% End of document
% --------------------------------------------------------------------------

\end{document}